\documentclass[letterpaper]{article}
\usepackage{aaai24}
\nocopyright
\usepackage{times}
\usepackage{helvet}
\usepackage{courier}
\usepackage[hyphens]{url}
\usepackage{graphicx}
\usepackage{natbib}
\usepackage{caption}
\usepackage{algorithm}
\usepackage{algorithmic}
\usepackage{newfloat}
\usepackage{listings}
\usepackage{multirow}
\usepackage{amsmath}

\DeclareCaptionStyle{ruled}{labelfont=normalfont,labelsep=colon,strut=off} 
\floatstyle{ruled}
\newfloat{listing}{tb}{lst}{}
\floatname{listing}{Listing}
\title{Nicer Than Humans: How do Large Language Models Behave in the Prisoner's Dilemma?}

\author {
    Nicoló Fontana\textsuperscript{\rm 1},
    Francesco Pierri\textsuperscript{\rm 1},
    Luca Maria Aiello\textsuperscript{\rm 2,3}
}
\affiliations {
    \textsuperscript{\rm 1}Politecnico di Milano, Italy\\
    \textsuperscript{\rm 2}IT University of Copenhagen, Denmark\\
    \textsuperscript{\rm 3}Pioneer Centre for AI, Denmark\\
    nicolo.fontana@mail.polimi.it, francesco.pierri@polimi.it, luai@itu.dk
}

\usepackage{enumitem}
\usepackage{xcolor}
\newcommand{\llamatwoname}[1]{\texttt{Llama-2-70b-chat-hf}}
\newcommand{\llamatwo}[1]{Llama2}
\newcommand{\llamathreename}[1]{\texttt{Llama-3-70B-Instruct}}
\newcommand{\llamathree}[1]{Llama3}
\newcommand{\gptname}[1]{\texttt{GPT-3.5-Turbo}}
\newcommand{\gpt}[1]{GPT3.5}
\newcommand{\gptthree}[1]{GPT3}
\newcommand{\gptfour}[1]{GPT4}
\newcommand{\ac}[1]{$\mathsf{AC}$}
\newcommand{\acname}[1]{$\mathsf{Always\ Cooperate}$} 
\newcommand{\ad}[1]{$\mathsf{AD}$} 
\newcommand{\adname}[1]{$\mathsf{Always\ Defect}$} 
\newcommand{\rnd}[1]{$\mathsf{RND}$} 
\newcommand{\rndname}[1]{$\mathsf{Random}$} 
\newcommand{\urnd}[1]{$\mathsf{URND}p$} 
\newcommand{\urndname}[1]{$\mathsf{Unfair\ Random}$} 
\newcommand{\tft}[1]{$\mathsf{TFT}$} 
\newcommand{\tftname}[1]{$\mathsf{Tit\ For\ Tat}$} 
\newcommand{\stft}[1]{$\mathsf{STFT}$} 
\newcommand{\stftname}[1]{$\mathsf{Suspicious\ Tit\ For\ Tat}$} 
\newcommand{\grim}[1]{$\mathsf{GRIM}$} 
\newcommand{\grimname}[1]{$\mathsf{Grim\ Trigger}$} 
\newcommand{\wsls}[1]{$\mathsf{WSLS}$}
\newcommand{\wslsname}[1]{$\mathsf{Win\mbox{--}Stay\ Lose\mbox{--}Shift}$} 
\newcommand{\cooperative}[1]{$\mathsf{Cooperative}$} 
\newcommand{\nice}[1]{$\mathsf{Nice}$} 
\newcommand{\forgiving}[1]{$\mathsf{Forgiving}$} 
\newcommand{\retaliatory}[1]{$\mathsf{Retaliatory}$} 
\newcommand{\troublemaking}[1]{$\mathsf{Troublemaking}$} \newcommand{\emulative}[1]{$\mathsf{Emulative}$} 
\newcommand{\consistent}[1]{$\mathsf{Consistent}$} 

\usepackage[section]{placeins}

\begin{document}
\maketitle

\begin{abstract}
The behavior of Large Language Models (LLMs) as artificial social agents is largely unexplored, and we still lack extensive evidence of how these agents react to simple social stimuli. Testing the behavior of AI agents in classic Game Theory experiments provides a promising theoretical framework for evaluating the norms and values of these agents in archetypal social situations. In this work, we investigate the cooperative behavior of three LLMs (\llamatwo{}, \llamathree{}, and \gpt{}) when playing the Iterated Prisoner's Dilemma against random adversaries displaying various levels of hostility. We introduce a systematic methodology to evaluate an LLM's comprehension of the game rules and its capability to parse historical gameplay logs for decision-making. We conducted simulations of games lasting for 100 rounds and analyzed the LLMs' decisions in terms of dimensions defined in the behavioral economics literature. We find that all models tend not to initiate defection but act cautiously, favoring cooperation over defection only when the opponent's defection rate is low. Overall, LLMs behave at least as cooperatively as the typical human player, although our results indicate some substantial differences among models. In particular, \llamatwo{} and \gpt{} are more cooperative than humans, and especially forgiving and non-retaliatory for opponent defection rates below 30\%. More similar to humans, \llamathree{} exhibits consistently uncooperative and exploitative behavior unless the opponent always cooperates. Our systematic approach to the study of LLMs in game theoretical scenarios is a step towards using these simulations to inform practices of LLM auditing and alignment.
\end{abstract}

\section{Introduction} \label{sec:intro}

Large Language Models (LLMs) can operate as social agents capable of complex, human-like interactions~\cite{Park-et-al_2023_GenerativeAgents}. Their integration into online social platforms is unfolding rapidly~\cite{cao2023comprehensive,yang2023anatomy}, presenting severe risks~\cite{floridi2020gpt,ferrara2024genai} as well as intriguing opportunities~\cite{dafoe2020open,Breum-et-al_2023_PersuasivePowerOfLLMs}. To understand and anticipate the behavioral dynamics that may arise from the interaction between artificial agents and humans, it is essential to first study how these agents react to simple social stimuli~\cite{bail2024generative}. Behavioral economics experiments, particularly those grounded in Game Theory, provide an ideal ground for testing the responses of AI agents to archetypal social situations~\cite{Horton_2023_LLMAsAsEconomicAgents}. These experiments typically involve goal-oriented scenarios where multiple `players' engage in a series of repeated interactions~\cite{osborne1994course}. To optimize for the goal, the decisions taken at each round must strategically account for the anticipated actions of the other players. However, the decisions of human participants often deviate from the theoretically optimal strategies due to the influence of social and psychological factors that conflict with the game's objectives~\cite{camerer97progress}. Similarly, given that LLMs are computational models built upon collective human knowledge and culture~\cite{schramowski2022large}, observing their behavior in classic iterated games could shed light on the social norms and values that these models reflect, as well as their capability in reasoning, planning, and collaborating in social settings.

Early interdisciplinary research has explored the use of LLMs within the context of classical economic games (see Related Work). While highly valuable, all these studies exhibit at least one of the following limitations. First, they generally lack prompt validation procedures, leading to an implicit assumption that LLMs can understand the complex rules of the game and the history of past actions described in the prompt~\cite{Akata-et-al_2023_RepeatedGamesWithLLMs, mao2023alympics, Mei-et-al_2024_TuringTestForAIChatbotsHumansSimilarity}. Second, the duration of simulated games is often limited to a few rounds~\cite{Akata-et-al_2023_RepeatedGamesWithLLMs, Brookins-et-al_2023_PlayingGamesWithGPT, Fan-et-al_2023_LLMsAsRationalPlayersInGT, Guo_2023_GPTInGTExperiments, Xu-et-al_2023_MAgIC}, hampering the LLMs' ability to discern the decision-making patterns of other participants -- a phenomenon we show in our own experiments. Third, the initialization of LLMs with predefined `personas' tends to skew their responses towards pre-determined behaviors, such as altruism or selfishness~\cite{Brookins-et-al_2023_PlayingGamesWithGPT, Fan-et-al_2023_LLMsAsRationalPlayersInGT, Guo_2023_GPTInGTExperiments, Horton_2023_LLMAsAsEconomicAgents, Lore-et-al_2023_StrategicBehaviorOfLLMs, phelps2023investigating}. This approach limits the exploration of the LLMs' baseline behavior, which is crucial for understanding their inherent decision-making processes. Last, the evaluation of simulation outcomes has predominantly focused on the quantitative analysis of decision types (e.g., frequency of cooperation), overlooking the LLMs' higher-level behavioral patterns that can be inferred from the temporal evolution of these decisions~\cite{Xu-et-al_2023_MAgIC, mao2023alympics}. The combined effect of these limitations has led to findings that are sometimes inconclusive~\cite{Brookins-et-al_2023_PlayingGamesWithGPT, mao2023alympics} and contradictory~\cite{Akata-et-al_2023_RepeatedGamesWithLLMs, Fan-et-al_2023_LLMsAsRationalPlayersInGT}, calling for more systematic evidence on the behavior of LLMs in iterated games.

In this work, we investigate the adaptability of LLMs in terms of their cooperative behavior when facing a spectrum of hostility in an iterated game scenario. We evaluate \llamatwo{}~\cite{touvron2023llama}, \llamathree{}, and \gpt{}~\citep{BrownTea20_LLMsAreFewShotLearners} in the Iterated Prisoner's Dilemma~\cite{osborne1994course} against adversaries with different propensities for betrayal. Our contribution is threefold. First, we introduce a meta-prompting technique designed to evaluate an LLM's comprehension of the game's rules and its ability to parse historical gameplay logs for decision-making. Second, we conduct extensive simulations over 100 rounds and determine the optimal memory span that enables the LLMs to adhere to the strategic framework of the game. Third, we analyze the behavioral patterns exhibited by the LLMs, aligning them with the core dimensions and strategies delineated in Robert Axelrod's influential research on the evolution of cooperation within strategic interactions~\cite{axelrod1981evolution}.

We observe that, overall, the three models tend to be more cooperative than humans, but they display some variability in their strategies. This variability persists even when the models are exposed to the same environment, game, and task framing. Both \llamatwo{} and \gpt{} displayed a more marked propensity towards cooperation than what existing literature reports about human players, indicating a favorable alignment with positive values.
In contrast, \llamathree{} adopts a more strategic and exploitative approach that is more similar to that of humans. This approach may be advantageous in competitive scenarios where raw performance is critical, but it can be a disadvantage when it comes to aligning with positive values.

Overall, our work contributes to defining a more principled approach to using LLMs for iterated games. It makes a step towards a more systematic way to use simulations of game theoretical scenarios to probe the inherent social biases of LLMs, which might prove useful for LLM auditing and alignment~\cite{shen2023large,mokander2023auditing}.

\section{Background on Prisoner's Dilemma} \label{sec:background}

\subsection{Game Setup} \label{sec:background:game}

The Prisoner's Dilemma is a classic thought experiment in Game Theory. It serves as a paradigm for analyzing conflict and cooperation between two players~\cite{Tucker-et-al_1983_TheMathematicsOfTucker}. In the game, the two players cannot communicate, and must independently choose between two actions: \emph{Cooperate}, or \emph{Defect}. Once both players have chosen their actions, payoffs are distributed based on the resulting combination of their choices. Mutual cooperation yields a reward $R$ for each player. If one defects while the other cooperates, the defector receives a higher `temptation' payoff $T$, while the cooperating player incurs a lower `sucker's' payoff $S$. If both parties choose to defect, they each receive a punishment payoff $P$ for failing to cooperate. The classical structure of the game is defined by the payoff hierarchy $T > R > P > S$, which theoretically incentivizes rational players to consistently choose defection as their dominant strategy~\cite{Axelrod_1981_EmergenceOfCoopAmongEgoists}. In the iterated version of the game, multiple rounds are played, and the payoffs are revealed to the players at every round~\cite{Tucker-et-al_1959_ContributionsToTheTheoryOfGames}. The iterative nature of the game allows the players to consider past outcomes to strategically inform future actions. When humans play the game, psychological factors such as reputation and trust significantly influence the decision-making process, often leading to higher rates of cooperation than would be expected from purely rational agents~\cite{DalBó-et-al_2011_EvoOfCoopInInfiniteIPD, Romero-et-al_2018_ConstructingStrategiesInIndefinitelyIPD}.

\subsection{Strategies} \label{sec:background:strategies}

In the Iterated Prisoner's Dilemma (IPD), a \emph{strategy} refers to an algorithm that a player uses to decide their next action, taking into account the historical context of the game~\cite{Kuhn_2019_StrategiesForTheIPD}. No single strategy universally outperforms all others; however, some are more effective against a broader variety of opposing strategies~\cite{DalBó-et-al_2011_EvoOfCoopInInfiniteIPD}. This concept was demonstrated in 1980 by Robert Axelrod~\cite{Axelrod_1980_EffectiveChoiceInPD}, who ran an IPD tournament with multiple competing strategies. Follow-up tournaments have further diversified the range of strategies~\cite{stewart2012extortion}.
Considering previous literature~\cite{Fudenberg-et-al_2012_SlowToAngerFastToForgive, DalBó-et-al_2011_EvoOfCoopInInfiniteIPD, Romero-et-al_2018_ConstructingStrategiesInIndefinitelyIPD}, we consider the strategies that better represent the ones adopted by humans, covering more than $75\%$ of the experimental samples in those studies. Those strategies are the following:
\begin{itemize}[leftmargin=*]
	\item \acname{} (\ac{}).
	\item \adname{} (\ad{}).
	\item \rndname{} (\rnd{}). Chooses Cooperate or Defect at random with equal probability at each round.
	\item \urndname{} (\urnd{}). Variation of Random where the probability of choosing to Cooperate is $p$.
	\item \tftname{} (\tft{}). Starts with Cooperation in the first round, then mimics the opponent's previous action throughout the game.
	\item \stftname{} (\stft{}). A \tft{} strategy that begins with Defect in the first round.
	\item \grimname{} (\grim{}). Chooses Cooperate until the opponent defects, then chooses only Defect for the rest of the game.
	\item \wslsname{} (\wsls{}). Repeats the previous action if it resulted in the highest payoffs ($R$ or $T$), otherwise changes action.
\end{itemize}
\tftname{} emerged as the winning strategy in Axelrod's tournament. It is commonly observed that human players tend to favor straightforward strategies such as \ad{}, \tft{}, or \grim{}~\cite{Romero-et-al_2018_ConstructingStrategiesInIndefinitelyIPD}. To describe the LLM behavior in terms of these known strategies, in our experiments we calculate the similarity of the LLM's game sequences with the sequences that these hardcoded strategies would generate when playing against the same opponent.

\subsection{Behavioral Dimensions} \label{sec:background:dimensions}

To identify the defining factors of different strategies, we combine dimensions already defined in prior studies that quantify salient behavioral properties of players based on the observed game sequences~\cite{Axelrod_1980_EffectiveChoiceInPD, Mei-et-al_2024_TuringTestForAIChatbotsHumansSimilarity}:
\begin{itemize}[leftmargin=*]
    \item \nice{}. Propensity to not be the first to defect. For a single instance of the Iterated Prisoner's Dilemma, it is defined as $1$ if the player is not the first to defect, $0$ otherwise.
    \item \forgiving{}. Propensity to cooperate again after an opponent's defection, defined as: $\frac{\#\mathsf{forgiven\_defection}}{\#\mathsf{opponent\_defection}+\#\mathsf{penalties}},$ where the number of penalties corresponds to the times that, after defecting, the opponent sought forgiveness by cooperating and the player did not forgive them, thus keeping defecting.
    \item \retaliatory{}. Propensity of defecting immediately after an opponent's uncalled defection, defined as: $\frac{\#\mathsf{reactions}}{\#\mathsf{provocations}}$
    \item \troublemaking{}. Propensity to defect unprovoked, defined as a counterpart of being retaliatory: $\frac{\#\mathsf{uncalled\_defection}}{\#\mathsf{occasions\_to\_provoke}}$, where an uncalled defection is a defection following a cooperation (or being the first action of the game) and an occasion to provoke is any cooperation from the opponent in the previous round.
    \item \emulative{}. Propensity to copy the opponent's last move: $\frac{\#\mathsf{mimic}}{N-1}$, where a \emph{mimic} occurs any time the player played the same action that the opponent played in the previous round and $N$ is the number of iterations of the game.
\end{itemize}
Strategies that are \nice{}, \forgiving{}, and \retaliatory{} (e.g., \tft{}) perform best against a wide variety of opponents. Human players tend to be particularly uncooperative when faced with games where the reward $R$ for cooperating is much lower than the temptation $T$ to betray the other player. In the indefinitely iterated version of the Prisoner's Dilemma with a fixed probability at every round for the game to terminate, human subjects tend to be more cooperative when the probability of ending the game is low. Usually, in games with a low chance of continuation and a big gap between $R$ and $T$, the majority of the strategies adopted are most similar to \adname{} (70\% to 90\%); in games with a longer potential time horizon and $R$ closer to $T$, humans tend to be more forgiving and \tftname{} explains a larger portion of the subjects' strategies~\cite{DalBó-et-al_2011_EvoOfCoopInInfiniteIPD}.

\section{Experimental Design} \label{sec:setup} 

\subsection{LLM Setup} \label{sec:setup:llm} 

In our experiments, we use \llamatwoname{} and \llamathreename{} as open-source language models developed by Meta and released under commercial use licenses\footnote{\url{https://ai.meta.com/llama/license/}}$^,$\footnote{\url{https://llama.meta.com/llama3/license/}}. We access these models through the Hugging Face platform, using their Inference API\footnote{\url{https://huggingface.co/inference-api/serverless}}. As a closed-source language model, we use \gptname{}, developed and hosted by OpenAI, accessed via their proprietary API\footnote{\url{https://openai.com/index/openai-api/}}. \gpt{} has been used in many early experiments on LLM agents in game-theoretical scenarios~\citep{Mei-et-al_2024_TuringTestForAIChatbotsHumansSimilarity, Lore-et-al_2023_StrategicBehaviorOfLLMs, Xu-et-al_2023_MAgIC}.

The models are initialized with a temperature value of $0.7$, that is equal to the default value for GPT models and is consistent with previous studies using models from the Llama family~\cite{Lore-et-al_2023_StrategicBehaviorOfLLMs, Xu-et-al_2023_MAgIC}. An analysis of robustness to different temperature settings ($0.1$ and $1.0$) is provided in the Appendix (Figure~\ref{fig:llama3_gpt_temperature_coop}).

We make the LLMs play a series of Iterated Prisoner's Dilemma games, each consisting of $N=100$ rounds. Due to the stochastic nature of the responses that LLMs generate, we repeat each game $k=100$ times and report the average results along with 95\% confidence intervals. To evaluate the models' adaptability to different degrees of environmental hostility, we repeat the experiment by matching them against \urnd{} opponents (defined in \S\ref{sec:background:strategies}) with varying probability of cooperation $\alpha\in[0,1]$. The final outcome of each game is a sequence containing pairs of binary values representing the actions of the LLM (player $A$) and the opponent (player $B$) at each round $i$:
\begin{equation}
G^\alpha_k = [(A_i, B_i)]_{i\in[1,N]}.
\end{equation}
From the $G^\alpha_k$ sequence, we extract the empirical probability of the LLM to cooperate at round $i$, calculated as the fraction of $i^{th}$ rounds in which the LLM cooperated over $k$ trials:
\begin{equation}
p^\alpha_{coop}(i) = \frac{1}{k}\sum_k{G^\alpha_k(A_i)}.
\end{equation}
We calculate the average cooperation probability throughout a game by averaging $p^\alpha_{coop}(i)$ over all $N$ rounds:
\begin{equation}
p^\alpha_{coop} = \frac{1}{N}\sum_{i=1}^{N} p^\alpha_{coop}(i).
\end{equation}

\subsection{Prompting} \label{sec:setup:prompting}

To implement the game, we design a fixed system prompt that outlines the game's \emph{rules}, including the payoff structure, and the player's \emph{objective} to \emph{`get the highest possible number of points in the long run'}. The variable part of the prompt includes the \emph{memory} of the game, namely a log of the history of the players' actions up to the current round, along with instructions for generating the next action. The complete prompt can be found in the Appendix (Figures~\ref{fig:system_prompt}, \ref{fig:contextual_prompt}, and \ref{fig:instructing_prompt}). 
In iterated games, the information from earlier rounds is essential for a player to deduce the opponent's strategy, and adapt accordingly. Early research involving LLMs in iterated games experimented with a limited number of rounds, precluding any analysis of how the size of the memory window influences the agent's behavior~\cite{Akata-et-al_2023_RepeatedGamesWithLLMs, Guo_2023_GPTInGTExperiments, Xu-et-al_2023_MAgIC}. 
In formulating the memory component, we explore various window sizes to provide the model with information from only the $n$ most recent rounds. We evaluate the effect of different memory window sizes by testing the LLM against an \adname{} opponent and identifying the optimal window size (see \S\ref{sec:results:window}). This assessment is based on the premise that, once the LLM has gathered sufficient information to recognize the opponent's consistently defecting behavior, its actions should align with defection, which is the only logical strategy.

\subsection{Meta-prompting} \label{sec:setup:metaprompting}

The development of effective LLM prompts is an ever-evolving practice. Although certain studies suggested guidelines for prompt development~\cite{ziems2305can}, achieving a consensus on the most effective prompting strategies across tasks remains challenging. Typically, the prompt quality is assessed empirically based on downstream performance~\cite{lester2021power}. This method is suitable for conventional classification or regression tasks where some form of ground truth is clearly defined. However, it is less applicable to generative tasks that lack a formal standard of correctness. In the specific context of the Prisoner's Dilemma, any sequence of Cooperate and Defect actions could be considered plausible. This ambiguity makes it difficult to discern whether LLM-generated sequences reflect a proper understanding of the game's rules or are merely the product of the model hallucinating~\cite{xu2024hallucination}. Prior research involving LLMs in Game Theory experiments has assessed outputs by requesting that the LLM provide a reasoned explanation of its output~\cite{Guo_2023_GPTInGTExperiments}. However, this approach offers only a retrospective justification, which can itself suffer from hallucinations if the LLM has not fully grasped the task's underlying instructions.

\begin{table}[t!]
\begin{tabular}{c p{1.5cm} p{5.3cm}}
 & \textbf{Name} & \multicolumn{1}{c}{\textbf{Question}} \\
\hline
\parbox[t]{1mm}{\multirow{3}{*}[-1.5em]{\rotatebox[origin=c]{90}{Rules}}}
& \texttt{min\_max} & What is the lowest/highest payoff player A can get in a single round? \\ 
& \texttt{actions} & Which actions is player A allowed to play? \\ 
& \texttt{payoff} & Which is player X's payoff in a single round if $X$ plays $p$ and $Y$ plays $q$? \\
\hline
\parbox[t]{2mm}{\multirow{3}{*}[-1.5em]{\rotatebox[origin=c]{90}{Time}}}
& \texttt{round} & Which is the current round of the game? \\
& \texttt{action$_i$} & Which action did player $X$ play in round $i$? \\
& \texttt{points$_i$} & How many points did player $X$ collect in round $i$? \\
\hline
\parbox[t]{2mm}{\multirow{3}{*}{\rotatebox[origin=c]{90}{State}}}
& \texttt{\#actions} & How many times did player $X$ choose $p$? \\
& \texttt{\#points} & What is player $X$'s current total payoff? \\
\end{tabular}
\caption{Templates of prompt comprehension questions used in meta-prompting to verify the LLM's comprehension of the prompt.}
\label{tab:meta-prompting}
\end{table}

To partially mitigate this issue, we introduce a novel meta-prompting technique to measure the LLMs' comprehension of a given prompt, to inform the process of prompt refinement. Specifically, we formulate a set of \emph{prompt comprehension} questions that address three key aspects of the prompt (see Table~\ref{tab:meta-prompting}): the \emph{game rules}, to verify the LLM's grasp of the game mechanics (e.g., \emph{`What is the lowest payoff that player A can get in a single round?'}); the chronological sequence of actions within the game history (e.g., \emph{`Which action did player A play in round 5?'}); and the cumulative game statistics (e.g., \emph{`What is player's $B$ current total payoff?'}).
To assess the LLMs' proficiency in responding to meta-prompting questions, we conduct a series of $3$ games of $100$ rounds each against \rnd{} opponents. We pose the questions at each round and compute the average accuracy of the LLMs' responses. At any given round $i$, a question template is instantiated into a set of questions that cover all possible combinations of the template's parameters. For example, at round $i$, questions referring to specific rounds (\texttt{action$_i$} and \texttt{points$_i$}) are asked for all past rounds from 1 to $i-1$.

\subsection{Behavioral Profiling} \label{sec:setup:behavioral}

We profile the LLMs' behavior with respect to a game history $G^\alpha_k$ in two ways. First, we quantify behavioral patterns by computing the behavioral dimensions outlined in \S\ref{sec:background:dimensions}. This computation results in a five-dimensional numerical vector that encapsulates the behavioral characteristics of the LLMs. Second, we use the Strategy Frequency Estimation Method (SFEM) defined in previous work~\cite{Romero-et-al_2018_ConstructingStrategiesInIndefinitelyIPD} to calculate the affinity between a player's game history and any of the classic strategies used in Prisoner Dilemma tournaments (see \S\ref{sec:background:strategies}). SFEM is a finite-mixture approach that uses likelihood maximization to estimate the likelihood of a strategy being represented in experimental data~\cite{Romero-et-al_2018_ConstructingStrategiesInIndefinitelyIPD}. Given a game history, SFEM outputs a score between $0$ and $1$ for each strategy in the set of theoretical strategies considered. Being a mixture approach, the sum of all the SFEM scores over the set of possible strategies does not have to sum up to $1$. The final SFEM scores we report are averaged over all the histories analyzed.

\section{Results} \label{sec:results}

\subsection{LLM's Prompt Comprehension}
\label{sec:results:comprehension}

\begin{figure}[t!]
 \centering 
 \includegraphics[width=\linewidth]{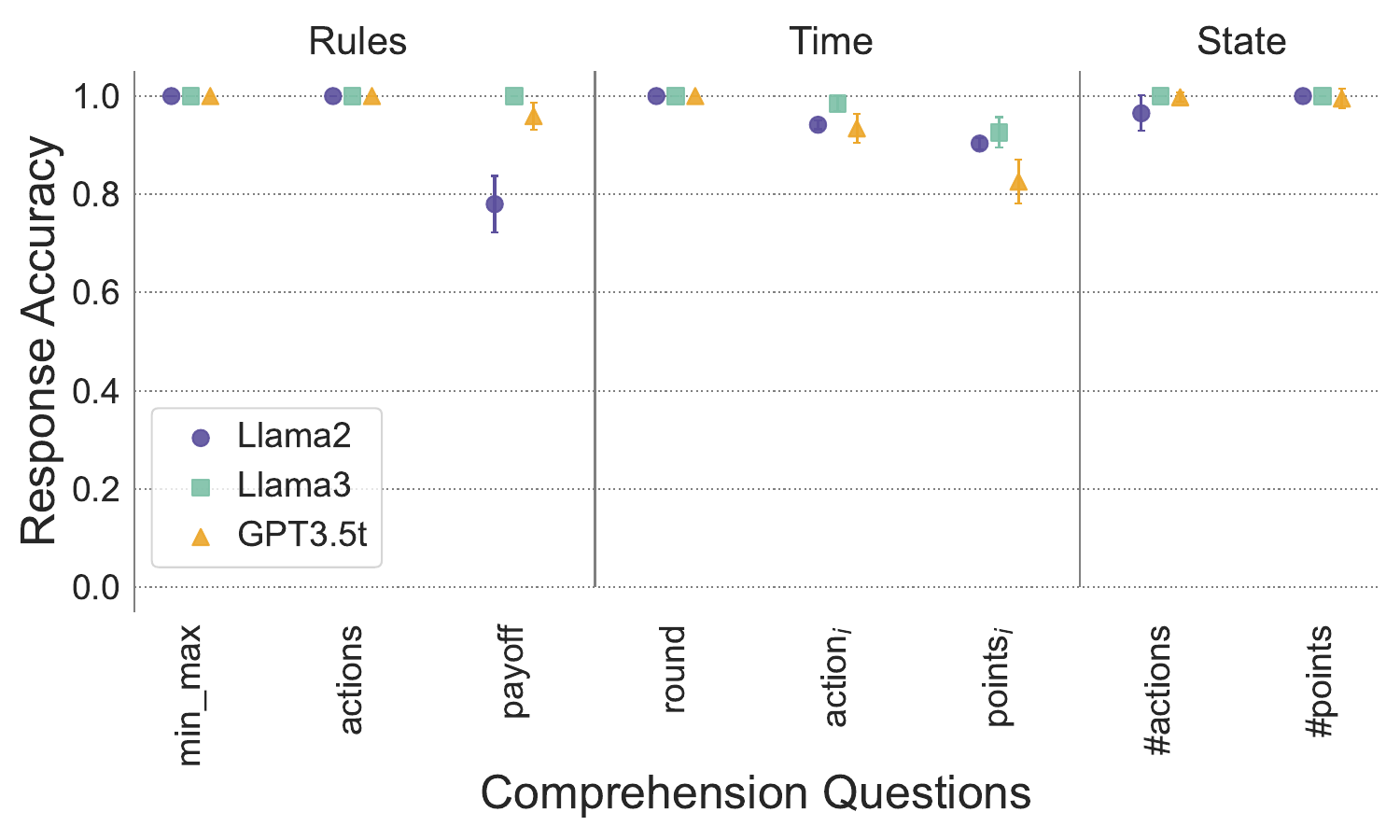} 
 \caption{Accuracy of the models' responses to the prompt comprehension questions defined in Table~\ref{tab:meta-prompting}. The questions are categorized into three groups, each assessing different aspects: the rules of the game, its temporal evolution, and its current state. We show 95\% confidence intervals, computed from 100 games.}
 \label{fig:questions_accuracy}
\end{figure}

Figure~\ref{fig:questions_accuracy} presents the accuracy of responses to the prompt comprehension questions associated with the final prompt that we used in our experiments, with results averaged over all trial runs. Overall, the models exhibit a good understanding of the concepts assessed by the questions, with most responses achieving an accuracy ranging from $0.8$ to $1.0$.

We iteratively tested multiple prompt versions against the response accuracy of \llamatwo{}. No further iterations of other models were needed, as the best prompt for \llamatwo{} yielded very high response accuracy in both \llamathree{} and \gpt{}.
Generally, adding explicit information about the game state and rules led to a better level of prompt comprehension. For illustration, Figure~\ref{fig:initial_prompt_questions_accuracy} in the Appendix includes a comparative analysis of accuracy scores derived from an earlier version of the prompt, which lacked a summary of the cumulative point totals for the players. In the absence of explicit total counts, \llamatwo{} was required to sum all points from the game history to determine the total tally, which significantly impacted its performance on the \texttt{\#actions} and \texttt{\#points} questions. This limitation aligns with previous findings that highlight that LLMs tend to struggle with arithmetic~\cite{Xu-et-al_2023_MAgIC, Aher-et-al_2023_UsingLLMsToSimulateHumans, Wei-et-al_2023_CoT}. The explicit inclusion of the sum of scores into the prompt markedly improved performance, achieving near-perfect accuracy.

\subsection{Effect of Memory Window Size} \label{sec:results:window}

\begin{figure}[t!]
 \centering 
 \includegraphics[width=\linewidth]{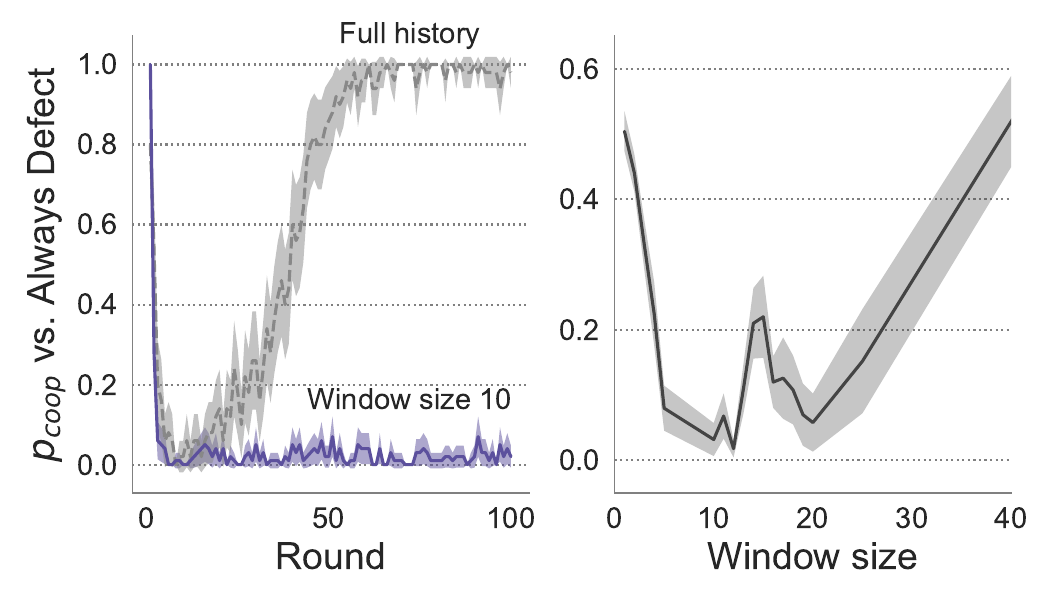} 
 \caption{\emph{Left}: Llama2's probability of cooperation ($p_{coop}$) against an \adname{} opponent, when using a memory window size of 10 vs. including the full game history in the prompt. \emph{Right}: steady-state probability calculated on the last 10 rounds of $p_{coop}$ for different memory windows sizes. We show 95\% confidence intervals, computed from 100 games.}  
 \label{fig:window_size_analysis}
\end{figure}

Figure~\ref{fig:window_size_analysis} (left) shows the probability of the LLM cooperating at each round in games of $100$ rounds, considering two memory window sizes: $m=100$ and $m=10$. Under both conditions, the LLM shifts towards a stance of consistent defection after approximately $5$ to $10$ rounds. Notably, this duration equals or exceeds the maximum number of rounds considered in previous studies. Without any constraints on the history length, the LLM's cooperation level quickly starts rising back, converging to full cooperation after the $50^{th}$ round. This pattern may be attributed to a combination of two factors: Llama2's intrinsic preference for positive constructs~\cite{Lore-et-al_2023_StrategicBehaviorOfLLMs} (favoring cooperation over defection) and its limited effectiveness in extracting actionable insights from long prompts~\cite{Xi-et-al_2023_RiseAndPotentialOfLLMAs}. With a memory window restricted to the $10$ most recent rounds, the LLM retains a full-defection stance throughout the game. We replicated this experiment across a range of memory window sizes and determined their respective equilibrium states by calculating the average cooperation probability in the final $10$ rounds (from the $90^{th}$ to the $100^{th}$). As illustrated in Figure~\ref{fig:window_size_analysis} (right), memory windows sizes around $10$ yield the expected outcome. Therefore, we select a window size of $m=10$ for the remainder of our experiments. 
Differently, \llamathree{} and \gpt{} show no variation in their cooperation levels regardless of the memory window size (see Figure~\ref{fig:llama3_gpt_window_size} in Appendix), allowing us to use the same prompt designed for \llamatwo{}. Notably, \llamathree{} does not suffer from the same bias towards cooperation as \llamatwo{}, and keeps its defective behavior also when provided with the full game history. \gpt{} exhibits the less optimal behavior, converging towards defection much more slowly.

\subsection{Behavioral Patterns} \label{sec:results:behavioral}

\begin{figure}[t!]
 \centering 
 \includegraphics[width=\linewidth]{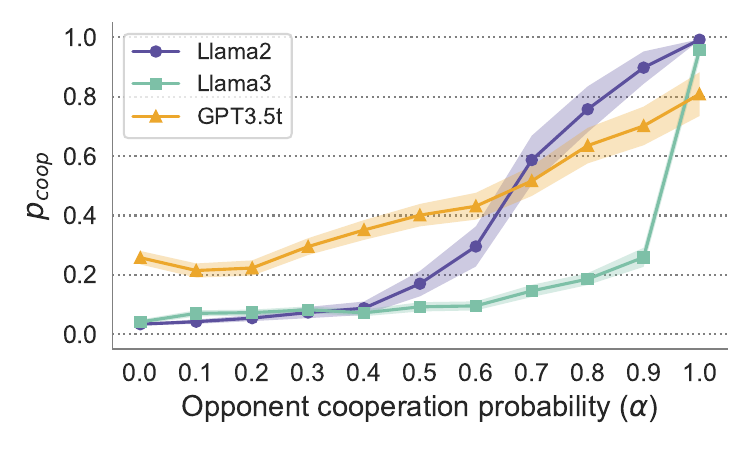} 
 \caption{Models' probability of cooperation ($p_{coop}$) against \urndname{} opponents with increasing cooperation probability $\alpha$. We show 95\% confidence intervals, computed from 100 games.}  
 \label{fig:pcoop_vs_urnd}
\end{figure}

\subsubsection{Probability of cooperation}
We examine the overall propensity for cooperation exhibited by the LLM across various degrees of environmental hostility. 
Figure~\ref{fig:pcoop_vs_urnd} shows the relationship between the probability of cooperative behavior $p^{\alpha}_{coop}$ of each model and the varying cooperation levels $\alpha$ of an \urndname{} opponent.

Among the three models, \llamathree{} demonstrates the most strategic approach. It maintains a very low level of cooperation even when the opponent is nearly always cooperating ($p_{coop}<0.3$ for every $\alpha<1$), but it increases its cooperation to nearly 100\% when the opponent is \ac{} ($\alpha=1$).
In contrast, \llamatwo{}'s behavior follows a sigmoidal curve across the entire range of $\alpha$ from $0$ to $1$. This indicates a rapid transition from a predominantly defecting strategy to a more cooperative attitude. The sigmoid curve is characterized by a relatively flat left tail, maintaining a stable probability of cooperation near $p_{coop}=0$ for $\alpha$ values up to $0.4$. The curve's inflection point is between $0.6$ and $0.7$—well beyond $0.5$—suggesting a cautious approach in interpreting the opponent's actions. Compared to \llamathree{}, the older Meta model is less guarded and more prone to increase its cooperation as soon as the opponent's cooperation significantly rises over the 50/50 chance.
\gpt{} displays a less strategic approach, maintaining a low but still significant cooperation level ($0.2<p_{coop}<0.4$) for low $\alpha$ ($<0.5$) and not reaching full cooperation even when the opponent is \ac{}.

In general, the models exhibit non-linear behaviors but with different characteristics. A common trend is the increasing cooperation as $\alpha$ grows, demonstrating a minimum level of strategic behavior for all LLMs.

These findings, especially concerning \llamathree{}, support previous research that highlights the cautious response patterns of LLMs in repeated game scenarios~\citep{Akata-et-al_2023_RepeatedGamesWithLLMs, phelps2023investigating}.

\subsubsection{SFEM profile}

The probability of cooperation provides a macroscopic perspective on a player's actions. However, to capture more nuanced strategic patterns that emerge during the match, we employed SFEM analysis (defined in \S\ref{sec:setup:behavioral}) to estimate the similarity between the behavioral patterns exhibited by the LLMs and those commonly observed in games involving human players.

\begin{figure}[t!]
 \centering 
 \includegraphics[width=\linewidth]{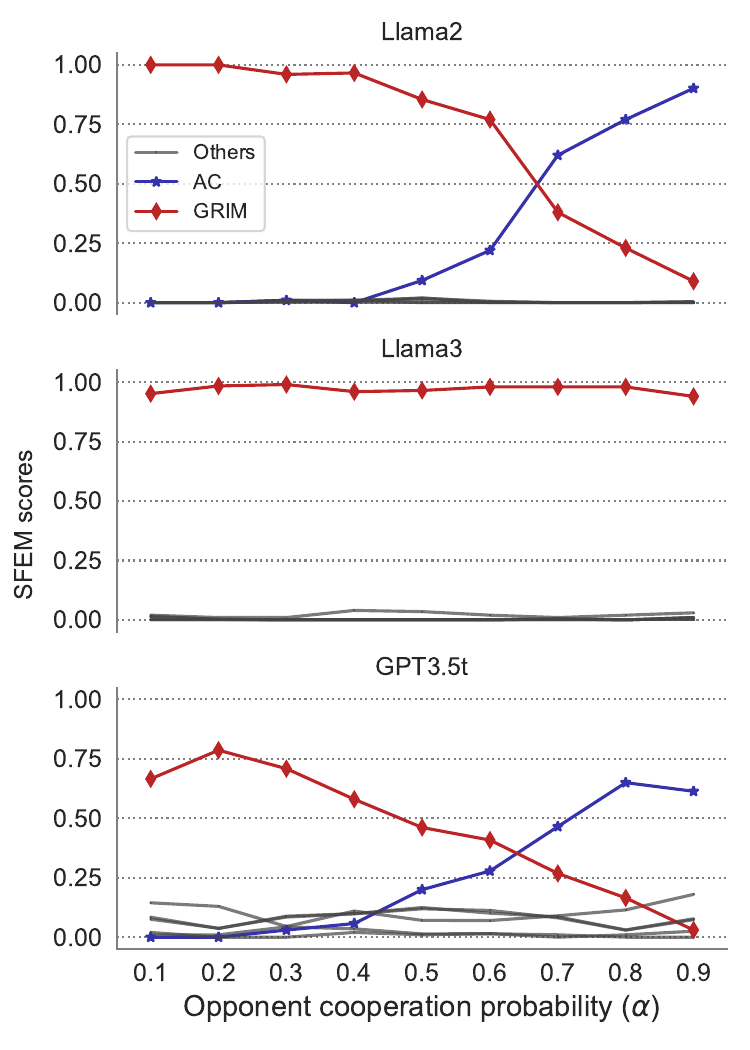} 
 \caption{SFEM scores quantifying the similarity between the models' sequences of actions and known strategies adopted in the Iterated Prisoner's Dilemma game (defined in \S\ref{sec:background:strategies}). The models' behavioral sequences come from games against \urndname{} opponents with increasing cooperation probability $\alpha$. Some SFEM scores are not shown because not well-defined for extreme values of $\alpha$.}
 \label{fig:sfem}
\end{figure}

Figure~\ref{fig:sfem} illustrates which strategies best explain the behaviors of each model as the values of $\alpha$ increase. When the adversary's probability of cooperation exceeds the $0.6$ threshold, there is a noticeable shift in both \llamatwo{}'s and \gpt{}'s strategy from \grimname{} to \acname{}.
In contrast, \llamathree{} consistently aligns with the \grimname{} strategy across all values of $\alpha$, displaying an exploitative approach even when the opponent tends to cooperate for the majority of the time. SFEM analysis further indicates that these strategies are the most representative of the LLMs' strategic behavior, with no other strategies being significantly indicative.

Previous studies  have shown that humans tend to be particularly uncooperative when the reward \( R \) is much lower than the temptation \( T \) and when playing for shorter time horizons, typically opting for startegies close to \adname{}{}~\citep{DalBó-et-al_2011_EvoOfCoopInInfiniteIPD, Romero-et-al_2018_ConstructingStrategiesInIndefinitelyIPD}. Conversely, humans become more cooperative when the reward is much higher and the time horizon is longer, opting for a mixture of \tftname{} and \grimname{} strategies.

Comparing this prior knowledge to our results indicates that \gpt{} and \llamatwo{} are consistently more cooperative than humans. In situations that do not favor cooperation—where the opponent defects more frequently than cooperates—they adopt the \grim{} strategy, unlike humans who tend to use the \ad{} strategy. Similarly, when the context supports cooperation (i.e., the opponent cooperates more than half of the time), both \gpt{} and \llamatwo{} employ the \ac{} strategy, which is notably more cooperative than the human preference for \tft{} or \grim{}. On the other hand, \llamathree{} is more cooperative than humans in hostile environments $(p_{\text{coop}} < 0.5$), also using the \grim{} strategy. However, in environments that encourage cooperation, \llamathree{} behaves more like humans by continuing to use the \grim{} strategy, whereas humans typically mix between the more forgiving \tft{} strategy and \grim{}.

\subsubsection{Behavioral profile}
\begin{figure}[t!]
 \centering 
 \includegraphics[width=\linewidth]{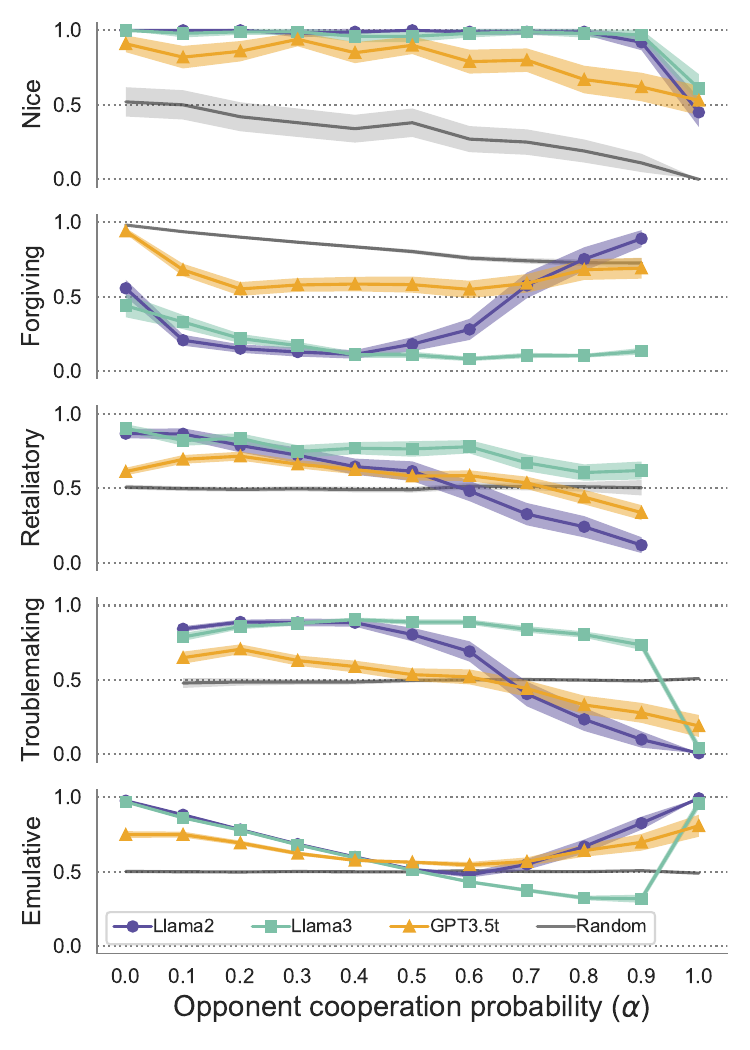} 
 \caption{Presence of behavioral traits in the models' actions when playing against \urndname{} opponents with increasing cooperation probability $\alpha$. The values of the same traits calculated for a \rndname{} agent playing against \urndname{} opponents are reported. Some traits are not defined for extreme values of $\alpha$. We show 95\% confidence intervals, computed from 100 games.}
 \label{fig:behavioral_profile}
\end{figure}

At a finer level of analysis, we characterize the behavior of the LLMs along the dimensions outlined in Section~\ref{sec:background:dimensions} (Figure~\ref{fig:behavioral_profile}).

When the parameter $\alpha$ is set to low values, the models exhibit highly uncooperative traits: they frequently retaliate following instances of betrayal, seldom revert to cooperative behavior after defecting, and are often the initiators of unprovoked defections. \gpt{} exhibits these traits in a more moderate manner, whereas \llamatwo{} and \llamathree{} display more extreme profiles in this regard.
More noticeable differences among the models' response emerge beyond $\alpha=0.5$. While \gpt{} and \llamatwo{} generate choices that are more \forgiving{} and less \troublemaking{}, \llamathree{} remains \troublemaking{} and rarely \forgiving{} for every $\alpha<1.0$.
On the other hand, the three models are constantly \nice{} across conditions, rarely defecting first. \gpt{} shows the lowest scores but remains above 0.5 even against an opponent that never defects ($\alpha=1$). In this scenario, all the LLMs never defect throughout the whole game in $50\%$ of the runs.
Interestingly, there is no scenario in which any LLM is simultaneously \nice{}, \forgiving{}, and \retaliatory{} --- the three conditions that characterize the most successful strategies, such as \tftname{}.

\section{Discussion and Conclusion}

Our study contributes to the broad literature on behavioral studies of Large Language Models as artificial social agents. Specifically, we study the responses of \llamatwo{}, \llamathree{}, and \gpt{} to the prototypical social scenario of the Prisoner's Dilemma. Building upon prior studies that explored the application of LLMs in classic Game Theory experiments, our work introduces a more systematic experimental setup that incorporates quantitative checks to better align the LLMs' responses to the complex task description. We have shown that aspects like prompt comprehension, memory representation, and duration of the simulation play crucial roles, especially for less powerful models, with the potential to significantly distort the experimental outcomes, if left unchecked. Our framework provides a quantitative guide for selecting the simulation variables and improving the prompt.

Our findings add a new benchmark to the body of work that explored the outcomes of iterated games, both among humans and among AI agents. In contrast to the behavioral patterns observed in humans playing the Prisoner's Dilemma game~\cite{DalBó-et-al_2011_EvoOfCoopInInfiniteIPD}, the three models displayed a more marked overall propensity towards cooperation. Under conditions that disincentivize cooperation, most human players adopt a stance of complete defection, whereas the LLMs' strategy, albeit mostly uncooperative, is characterized by an initial trust in the opponent's cooperation (\nice{}), reminiscent of the strategy known in Game Theory as \grimname{}. When the environment is more favorable to cooperative play, \llamathree{} playing \grimname{} becomes comparable to human strategies that often resemble either \grimname{} or \tftname{}. Differently, \llamatwo{} and \gpt{} tend towards a consistently cooperative approach. Notably, \llamatwo{}'s shift from \grimname{} to \acname{} occurs quite abruptly as the opponent's defection probability drops below 30\%, while \gpt{}'s transition is smoother and starts when $\alpha$ exceeds $0.2$.

Our results are in line with early experiments involving LLMs, which indicated a tendency for these models to cooperate in repeated games~\cite{Brookins-et-al_2023_PlayingGamesWithGPT, Mei-et-al_2024_TuringTestForAIChatbotsHumansSimilarity}. However, the broader research in this area has yielded mixed outcomes~\cite{Akata-et-al_2023_RepeatedGamesWithLLMs, phelps2023investigating}. Distinct from previous studies, our findings are derived from extensive game simulations conducted over numerous rounds and benefit from an experimental framework that leverages quantitative checks for accuracy.

Overall, our findings offer a robust baseline for understanding LLMs' behaviors in the Iterated Prisoner's Dilemma (IPD), a widely used Game Theory experiment for assessing agents' responses to conflictual scenarios. Establishing this baseline allows for clearer differentiation between the inherent tendencies of the models and the effects of specific elements within the game setting.

However, it is important to acknowledge the limitations of our work, which open the way for further research. First, our analysis was conducted using three models, specifically \llamatwo{}, \llamathree{}, and \gpt{}, which, at the time of writing, are among the most advanced models available~\cite{touvron2023llama, BrownTea20_LLMsAreFewShotLearners}\footnote{\url{https://ai.meta.com/blog/meta-llama-3/}}. Nevertheless, the field is progressing at an unprecedented pace, with new models being introduced regularly. To determine whether the behavioral patterns observed in our study are consistent across many different models, it is important to conduct comparative analyses with models that vary significantly in terms of parameter size and the volume of their training data. Second, our study's scope was limited to assessing the LLM's responses to \emph{random} strategies, and with a fixed payoff structure. Exploring the LLM's interactions with more sophisticated opponents would enable us to better delineate the boundaries of LLMs' inferential abilities in social contexts, and to draw more detailed behavioral profiles under a broader spectrum of conditions that more comprehensively represent prototypical social scenarios. We explored only the use of Zero-Shot Chain-of-Thought techniques without obtaining any improvement. Other options like Tree-of-Thought~\citep{YaoSea23_TreeOfThoughts} techniques or external modules~\citep{WangLea24_SurveyLargeLanguage} can be employed to test different reasoning conditions for the models.
Furthermore, the experimental framework of our study considers only a single LLM agent. Creating social groups of AI agents that interact through iterated games like the Prisoner's Dilemma would open up a wealth of opportunities to study emergent behaviors in synthetic societies, an avenue of research that is increasingly recognized as fundamental for a proper understanding of how LLMs can affect human societies~\cite{bail2024generative}. Last, despite the numerical guidelines we implemented to evaluate the quality of the prompt, our refinements of the prompt were not guided by any principle other than experience and instinct. Our attempt to use Chain-of-Thought~\cite{Kojima-et-al_2022_ZSCoT, Zhou-et-al_2023_APE-AutoZSCoT} as a structured way to approach prompt revision resulted in prompts that did not improve performance in our prompt comprehension question inventory (see Figure~\ref{fig:zscot_questions_accuracy} in Appendix). In this respect, our work provides yet another example of how prompt engineering would benefit from supporting tools to constrain its highly discretionary nature. Another interesting direction for future research is to explore persona prompting~\cite{hu2024quantifying}, where models are guided towards specific behaviors, such as altruism or selfishness. However, in this study, we intentionally avoided using personas to influence the outputs of the LLMs, as our primary objective is to assess the models' \emph{inherent biases} towards different behavioral patterns in game-theoretical scenarios. Similarly, models could be enhanced by incorporating a planning module that considers the potential outcomes of various actions over multiple steps~\cite{kambhampati2024llms}.

Despite its limitations, our work has two main implications. From the theoretical perspective, it expands our knowledge of the inherent biases of LLMs in social situations, which is crucial to inform their deployment across different contexts. From the practical perspective, it provides a principled way to approach game theoretical simulations with LLMs. This constitutes a step towards using these simulations as reliable and reproducible tools that could be used as tests to verify LLM alignment to desired principles of social cooperation~\cite{shen2023large}.

\section{Related work}

Next, we briefly review previous work using LLMs for social reasoning, the generation of human-like synthetic data, and simulations of human behavior.

\subsection{LLMs as Agents}

\citet{Argyle-et-al_2023_OutOfOneMany-LLMsToSimHumans} instructed LLMs to answer surveys as if they belonged to specified socio-demographic groups. They showed a high similarity between the responses generated by the LLM and those provided by the demographic groups it was asked to emulate. LLMs impersonating human agents with different profiles were used to explore the negotiation abilities of the models~\citep{DavidsonTea23_EvaluatingLanguageModel} or to create synthetic social networks, to observe emergent social behavior, most notably opinion dynamics and information spreading~\citep{Chuang-et-al_2023_OpinionDynInLLMANet, DeMarzo-et-al_2023_EmergenceOfSFNetsAmongLLMsInteractions, He-et-al_2023_HomophilyLLMAsInChirperOSN}. \citet{Park-et-al_2023_GenerativeAgents} developed a society with synthetic agents interacting with elements of a synthetic world, showed that those agents were able to adopt behaviors that are typical of humans without being directly prompted to do so. Using the same framework, \citet{Ren-et-al_2024_EmergenceOfSocialNormsInLLMASociety} showed that those agents were also able to build and spread social norms, while~\citet{PiattiGea24p_CooperateOrCollapse} developed a similar framework to investigate the robustness of LLMs societies.

\subsection{LLMs in Game Theory} 

Early work on the application of LLMs to Game Theory experiments touched upon both 1-time and iterated games.
Single-iteration experiments offer limited insight into LLM behavior. \citet{Brookins-et-al_2023_PlayingGamesWithGPT} showed that LLMs are more biased towards fairness and cooperation when compared to a human baseline sample. In contrast, \citet{Aher-et-al_2023_UsingLLMsToSimulateHumans} found that there is an overall alignment between the LLM-based agent behavior and the human ones. Investigating the ability of LLMs to predict human choices in 1-time games, \citet{CapraroVea24p_AssessingLargeLanguage} showed that only more powerful models are capable of doing it, although they overestimate the altruistic tendency of human players. 
When focusing on iterated games, the spectrum of patterns that can be identified expands, allowing more refined analysis. For example, \citet{Akata-et-al_2023_RepeatedGamesWithLLMs} managed to identify that LLMs can be particularly unforgiving. \citet{Mei-et-al_2024_TuringTestForAIChatbotsHumansSimilarity} discovered instead that the same models show a higher cooperation rate than compared to humans. \citet{Fan-et-al_2023_LLMsAsRationalPlayersInGT} exploited the iteration of games to check the level of the opponent's strategy that the LLM was able to infer from the history of actions. They showed that the inference capability of the LLM is limited, calling for more systematic approaches to structure memory and prompts. Testing a novel benchmark with different games, \citet{DuanJea24p_GTBench} studied multiple LLMs showing that closed-source models tend to achieve better performance than open-source ones. Although they did not examine the behavioral characteristics of the LLMs' responses, they also found that Zero-Shot Chain-of-Thought techniques are not always beneficial, which aligns with our results.

\section*{Ethical Considerations}

The deployment of Large Language Models as AI social agents raises numerous ethical considerations that are currently the subject of intense scrutiny by the interdisciplinary research community. The extraordinary capabilities of these models to generate text have led several scientists to envision alarming scenarios in which the seamless integration of AI agents into the online social discourse may facilitate the dissemination of harmful content, the spread of misinformation, and the propagation of `semantic garbage', ultimately damaging our societies~\cite{floridi2020gpt,weidinger2022taxonomy,hendrycks2023overview}. As a result, any research exploring the characteristics of LLMs as social agents could, directly or indirectly, contribute knowledge that might be exploited to implement and deploy LLM-based technologies for malicious purposes. While recognizing this risk, we also believe that conducting research on LLM-based agents is essential to assess potential risks and to guide efforts aimed at developing strategies to mitigate them. Our study contributes positively to deepen our understanding of how LLMs react to social stimuli.

Even when deploying LLM-based agents for ethical purposes, trade-offs between the obtained benefit and the high level of power consumption required to run them should be carefully considered~\cite{bender2021dangers}.

\section*{Code and Data Availability}

All code, prompts, and game traces will be made available on GitHub.

\section*{Acknowledgments}
This work was partially supported by the Italian Ministry of Education ( PRIN grant DEMON prot. 2022BAXSPY) and the European Union (NextGenerationEU project PNRR-PE-AI FAIR).
NF acknowledges the support from the Danish Data Science Academy through the DDSA Visit Grant (Grant ID: 2023-1856) and from Politecnico di Milano through the scholarship ``Tesi all'estero a.a. 2023/24-Primo bando''. LMA acknowledges the support from the Carlsberg Foundation through the COCOONS project (CF21-0432).

\clearpage

\appendix

\setcounter{footnote}{0} 
\setcounter{figure}{0}
\setcounter{table}{0}
\setcounter{equation}{0}
\renewcommand{\thefigure}{A\arabic{figure}}
\renewcommand{\thetable}{A\arabic{table}}
\renewcommand{\theequation}{A\arabic{equation}}

\section*{Appendix}\label{sec:appendix}

\subsection*{Prompts and their Variations}

\begin{figure}[!t]
 \centering 
 \includegraphics[width=\linewidth]{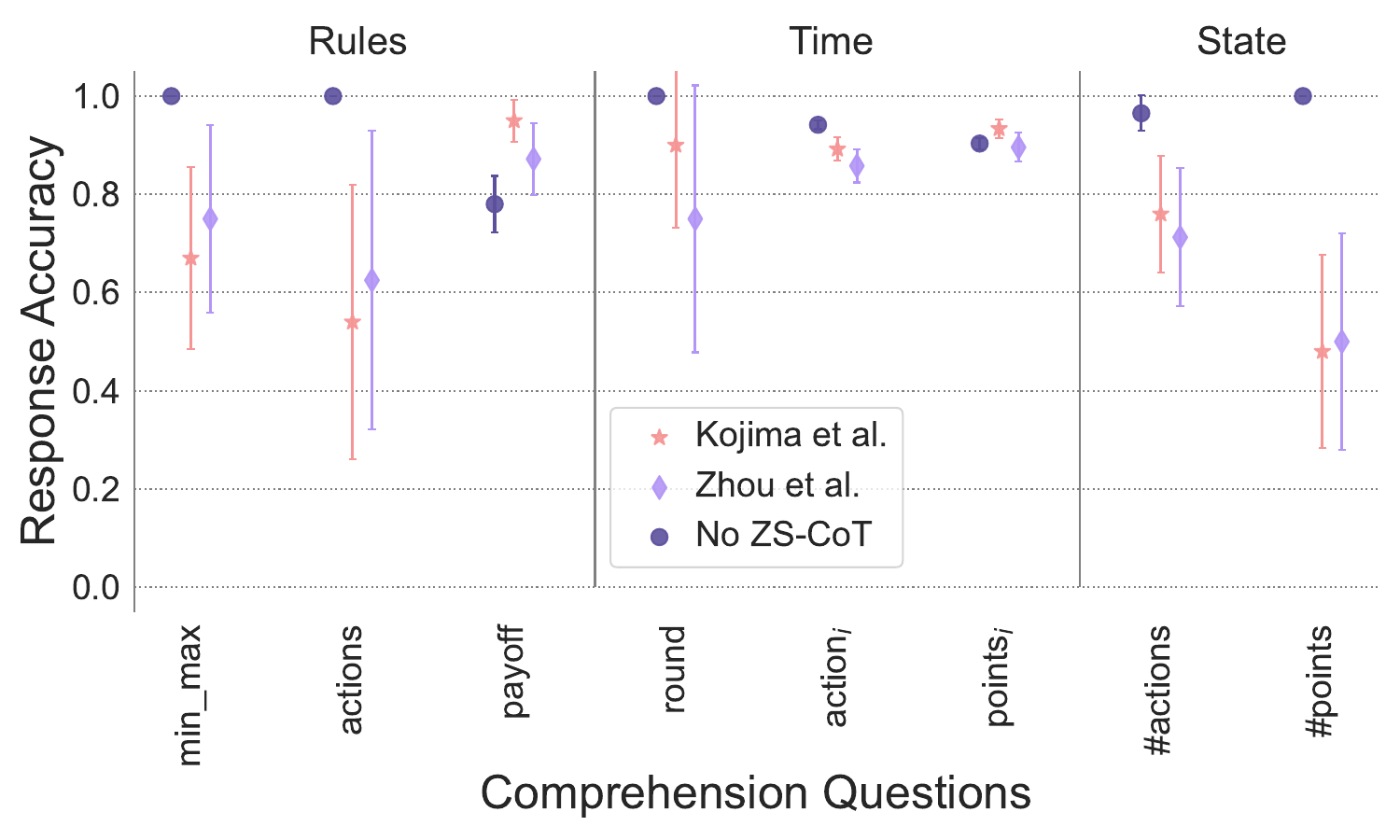} 
 \caption{Accuracy of \llamatwo{}'s responses to the prompt comprehension questions defined in Table~\ref{tab:meta-prompting} using two Zero-Shot Chain-of-Thought variations of the prompt compared to the accuracy obtained with the unvaried prompt. We show 95\% confidence intervals, computed from 100 games.}
 \label{fig:zscot_questions_accuracy}
\end{figure}

Figures~\ref{fig:system_prompt}, \ref{fig:contextual_prompt} and \ref{fig:instructing_prompt} show examples of the three parts of the final prompt we used in our experiments. Figures~\ref{fig:instructing_prompt_kojima} and \ref{fig:instructing_prompt_zhou} respectively show the variation in the instructing prompt when using the Zero-Shot Chain-of-Thought techniques from \citet{Kojima-et-al_2022_ZSCoT} and \citet{Zhou-et-al_2023_APE-AutoZSCoT}. Figure~\ref{fig:zscot_questions_accuracy} displays the accuracy obtain by \llamatwo{} in the comprehension questions when using the two variations and the unvaried prompt.

\begin{figure}[!t]
 \centering 
 \includegraphics[width=\linewidth]{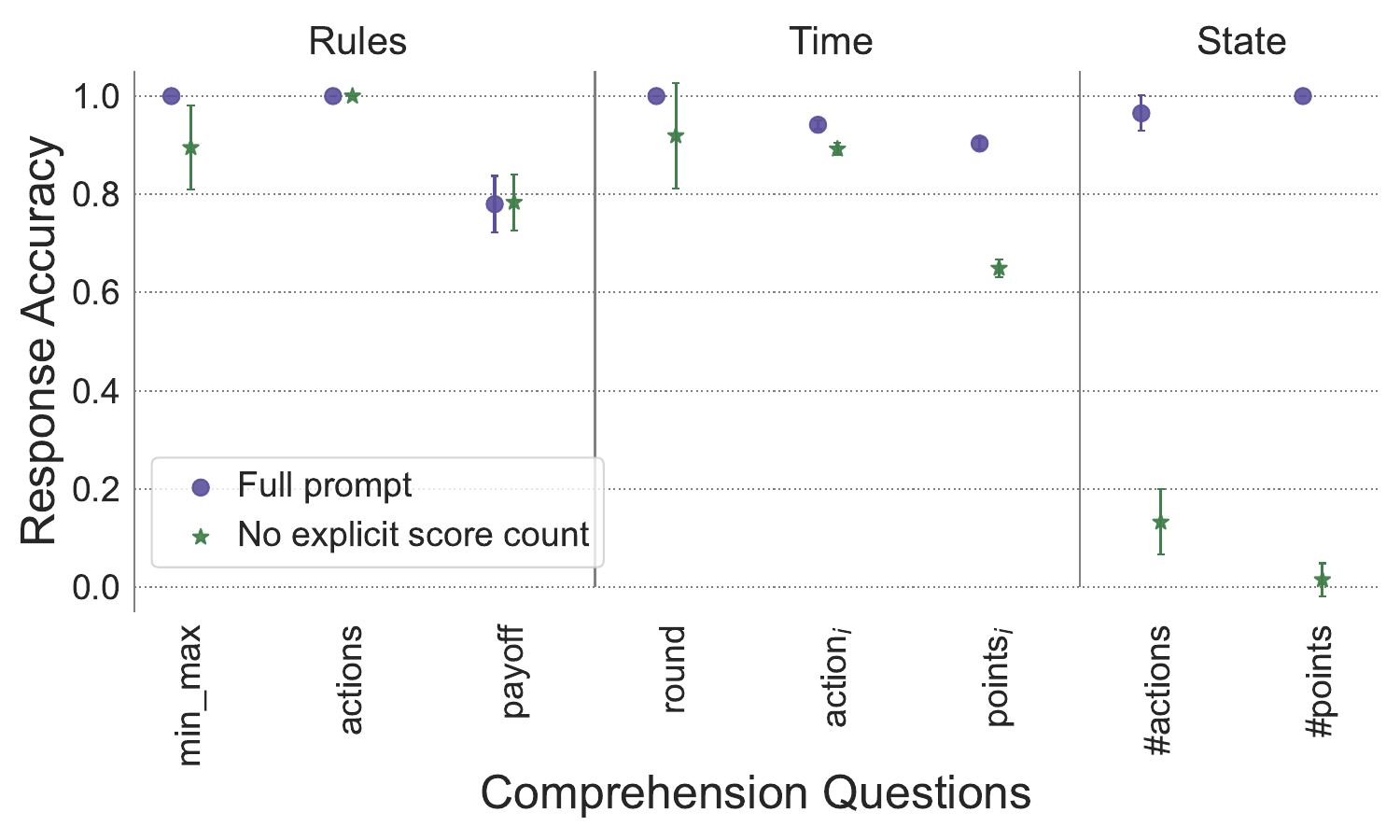} 
 \caption{Accuracy of \llamatwo{} responses to the prompt comprehension questions defined in Table~\ref{tab:meta-prompting}. The accuracy of the final full prompt is compared with an earlier version of the prompt that lacked a summary of the cumulative rewards achieved by the players. We show 95\% confidence intervals, computed from 100 games.}  
 \label{fig:initial_prompt_questions_accuracy}
\end{figure}

Figure~\ref{fig:initial_prompt_questions_accuracy} shows the comparison in the accuracy of \llamatwo{} between the initial prompt without explicit score counts and the final version of it.

\begin{figure}[!h]
    \small
    \resizebox{\columnwidth}{!}{
    \fbox{\begin{tabular}{p{\columnwidth}}
    \textbf{System prompt} \\
    \hline
    \texttt{\textless s\textgreater [INST]}
    
    \texttt{\textless\textless SYS\textgreater\textgreater}
    \vskip 1\baselineskip
    
    \texttt{Context: Player A and player B are playing a multi-round game.}
    
    \texttt{At each turn player A and player B simultaneously perform one of the following actions: \{"Cooperate", "Defect"\}}
    \vskip 1\baselineskip
    
    \texttt{The payoffs for each combination of chosen actions are the following:}
    
    \texttt{If A plays "Defect" and B plays "Defect", A collects 1 points and B collects 1 points.}
    
    \texttt{If A plays "Defect" and B plays "Cooperate", A collects 5 points and B collects 0 points.}
    
    \texttt{If A plays "Cooperate" and B plays "Defect", A collects 0 points and B collects 5 points.}
    
    \texttt{If A plays "Cooperate" and B plays "Cooperate", A collects 3 points and B collects 3 points.}
    \vskip 1\baselineskip
    
    \texttt{They will play a total of 100 rounds of this game.}
    \vskip 1\baselineskip
    
    \texttt{Remember that a player's objective is to get the highest possible number of points in the long run.}
    \vskip 1\baselineskip
    
    \texttt{\textless\textless SYS\textgreater\textgreater}
    \end{tabular}}}
    \caption{\emph{System prompt} describing the players' roles, the rules of the games, the number of rounds to be played, and the player's objective. This part of the prompt is fixed in all simulations.}
    \label{fig:system_prompt}
\end{figure}
\FloatBarrier

\begin{figure}[!t]
    \small
    \resizebox{\columnwidth}{!}{
    \fbox{\begin{tabular}{p{\columnwidth}}
    \textbf{Contextual prompt} \\
    \hline
    \vskip 1\baselineskip
    \texttt{The history of the game in the last 5 rounds is the following:}
    \vskip 1\baselineskip
    
    \texttt{Round 2: A played "Cooperate" and B played "Defect" A collected 0 points and B collected 5 points.}
    
    \texttt{Round 3: A played "Defect" and B played "Defect" A collected 1 points and B collected 1 points.}
    
    \texttt{Round 4: A played "Cooperate" and B played "Defect" A collected 0 points and B collected 5 points.}
    
    \texttt{Round 5: A played "Defect" and B played "Cooperate" A collected 5 points and B collected 0 points.}
    
    \texttt{Round 6: A played "Defect" and B played "Defect" A collected 1 points and B collected 1 points.}
    \vskip 1\baselineskip
    
    \texttt{In total, A chose "Cooperate" 2 times and chose "Defect" 3 times, B chose "Cooperate" 1 times and chose "Defect" 4 times.}
    
    \texttt{In total, A collected 7 points and B collected 12 points.}
    \vskip 1\baselineskip
    
    \texttt{Current round: 7.}
    \vskip 1\baselineskip
    \end{tabular}}}
    \caption{\emph{Contextual prompt} containing information about: the game history in the last $n$ rounds ($5$ in this example), the overall amount of times each player chose each action, the overall amount of points collected by each player, and the current round. For each turn, the prompt contains: the action played by each player and the points collected by each player. This prompt changes at each round.}
    \label{fig:contextual_prompt}
\end{figure}
\FloatBarrier
 
\begin{figure}[!t]
    \small
    \resizebox{\columnwidth}{!}{
    \fbox{\begin{tabular}{p{\columnwidth}}
    \textbf{Instructing prompt} \\
    \hline
    \vskip 1\baselineskip
    \texttt{Remember to use only the following JSON format:}
    
    \texttt{\{"action": \textless ACTION\_of\_A\textgreater, "reason": \textless YOUR\_REASON\textgreater\}}
    \vskip 1\baselineskip
    
    \texttt{Answer saying which action player A should play.}
    \vskip 1\baselineskip
    
    \texttt{Remember to answer using the right format.[/INST]}
    
    \vskip 1\baselineskip
    \end{tabular}}}
    \caption{\emph{Instructing prompt}. The LLM is instructed on the nature and format of the answer. This part of the prompt component is replaced with prompt comprehension questions in the phase of prompt tuning.}
    \label{fig:instructing_prompt}
\end{figure}
\FloatBarrier

\begin{figure}[!t]
    \small
    \resizebox{\columnwidth}{!}{
    \fbox{\begin{tabular}{p{\columnwidth}}
    \textbf{Instructing prompt} \\
    \hline
    \vskip 1\baselineskip
    \texttt{Remember to use only the following JSON format:}
    
    \texttt{\{"action": \textless ACTION\_of\_A\textgreater, "reason": \textless YOUR\_REASON\textgreater\}}
    
    \texttt{Answer saying which action player A should play.}
    
    \texttt{Remember to answer using the right format.[/INST]}
    
    \vskip 1\baselineskip

    \texttt{Let’s think step by step}
    
    \vskip 1\baselineskip
    \end{tabular}}}
    \caption{Variation of the \emph{instructing prompt} with \citet{Kojima-et-al_2022_ZSCoT} Zero-Shot Chain-of-Thought.}
    \label{fig:instructing_prompt_kojima}
\end{figure}
\FloatBarrier

\begin{figure}[!t]
    \small
    \resizebox{\columnwidth}{!}{
    \fbox{\begin{tabular}{p{\columnwidth}}
    \textbf{Instructing prompt} \\
    \hline
    \vskip 1\baselineskip
    \texttt{Remember to use only the following JSON format:}
    
    \texttt{\{"action": \textless ACTION\_of\_A\textgreater, "reason": \textless YOUR\_REASON\textgreater\}}
    
    \texttt{Answer saying which action player A should play.}
    
    \texttt{Remember to answer using the right format.[/INST]}
    
    \vskip 1\baselineskip

    \texttt{Let's work this out in a step-by-step way to be sure we have the right answer}
    
    \vskip 1\baselineskip
    \end{tabular}}}
    \caption{Variation of the \emph{instructing prompt} with \citet{Zhou-et-al_2023_APE-AutoZSCoT} Zero-Shot Chain-of-Thought.}
    \label{fig:instructing_prompt_zhou}
\end{figure}
\FloatBarrier

\FloatBarrier

\subsection*{Effect of Memory Window Size}
\begin{figure}[t!]
 \centering 
 \includegraphics[width=\linewidth]{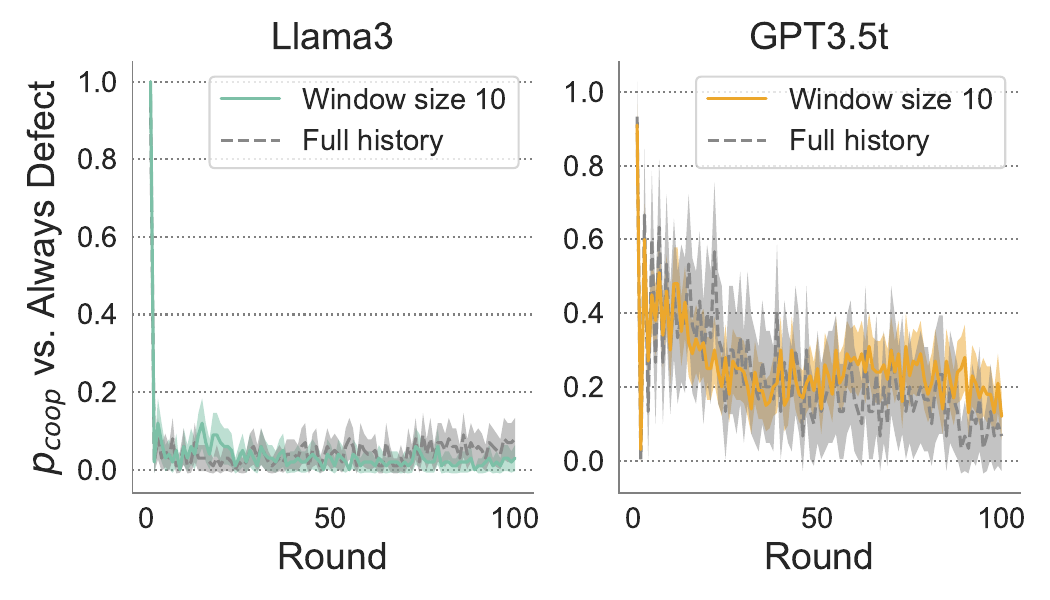} 
 \caption{\llamathree{}'s and \gpt{}’s probability of cooperation ($p_{coop}$)
against an Always Defect opponent, when using a memory
window size of 10 vs. including the full game history in the
prompt.}  
 \label{fig:llama3_gpt_window_size}
\end{figure}

Figure~\ref{fig:llama3_gpt_window_size} shows the probability of cooperation for \llamathree{} and \gpt{} when using memory windows of different sizes. Different from \llamatwo{}, neither of these models significantly changes its behavior depending on the window size. In particular, the trend observed for \gpt{}, which tends to stabilize around the value 0.2, is in accordance with the average $p_{coop}$ shown in Figure~\ref{fig:behavioral_profile} for $\alpha=0$

\FloatBarrier

\clearpage

\subsection*{Effect of Temperature}
\begin{figure}[t!]
 \centering 
 \includegraphics[width=\linewidth]{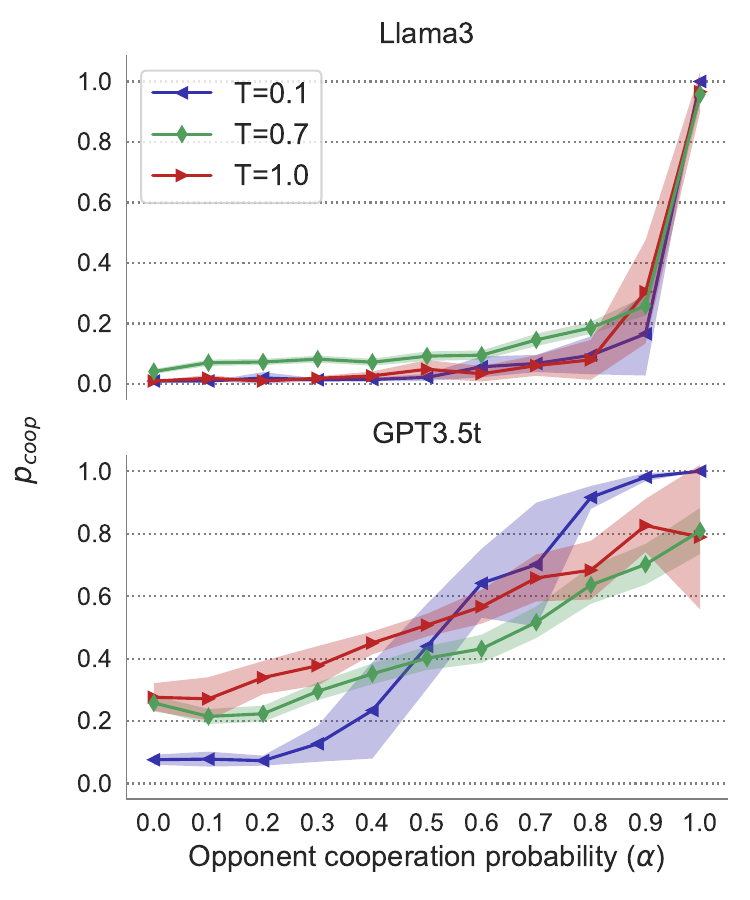} 
 \caption{\llamathree{}'s and \gpt{}’s probability of cooperation ($p_{coop}$) against \urndname{} opponents with increasing cooperation probability $\alpha$ for different values of the temperature hyperparameter. We show 95\% confidence intervals, computed from 10 games.}  
 \label{fig:llama3_gpt_temperature_coop}
\end{figure}

We explore the impact of varying the temperature hyperparameter on the probability of cooperation $p_{coop}$ of \llamathree{} and \gpt{}\footnote{HuggingFace discontinued access to \llamatwo{} through their Inference API in May 2024. Since then, the model has been available only dedicated server hosting at considerable costs. We therefore limit our tests on temperature to \llamathree{} and \gpt{} only.}. For each model and each temperature ($0.1$ and $1.0$), we run 10 games of 100 rounds each against opponents with varying cooperation probability $\alpha$.
Figure~\ref{fig:llama3_gpt_temperature_coop} presents the average probability of cooperation for \llamathree{} and \gpt{} for three different temperature values up to $1.0$. At temperatures greater than $1.0$, the models tend to produce seemingly random tokens, which renders their output unusable.

The Pearson correlation between the values of $p_{coop}$ across different temperature pairs is in the range $[0.97-1]$, indicating that the temperature does not affect the general trends of $p_{coop}$ as $\alpha$ varies. However, the growth of the cooperation curve for \gpt{} turns from roughly linear to a sigmoid as temperature decreases. This suggests that, in some models, different levels of determinism can influence the boundaries of their decision states. In this specific case, it appears that the noise that the increased temperature adds to the generated choices smoothens the sharp transition between the uncooperative and cooperative states that can be observed at low temperatures.

\clearpage

\end{document}